# Dynamic and Rapid Deep Synthesis of Molecular MRI Signals

Dinor Nagar[1], Nikita Vladimirov[2], Christian T. Farrar[3], Or Perlman[2,4*]

[1]*School of Electrical Engineering, Tel Aviv University, Tel Aviv, Israel*

[2]*Department of Biomedical Engineering, Tel Aviv University, Tel Aviv, Israel*

[3]*Athinoula A. Martinos Center for Biomedical Imaging, Department of Radiology, Massachusetts General Hospital and Harvard Medical School, Charlestown, MA, USA*

[4]*Sagol School of Neuroscience, Tel Aviv University, Tel Aviv, Israel*

*Correspondence to: Or Perlman, Department of Biomedical Engineering and Sagol School of Neuroscience, Tel Aviv University, Tel Aviv 6997801, Israel. Email: orperlman@tauex.tau.ac.il

**Abstract**

Model-driven analysis of biophysical phenomena is gaining increased attention and utility for medical imaging applications. In magnetic resonance imaging (MRI), the availability of well-established models for describing the relations between the nuclear magnetization, tissue properties, and the externally applied magnetic fields has enabled the prediction of image contrast and served as a powerful tool for designing the imaging protocols that are now routinely used in the clinic. Recently, various advanced imaging techniques have relied on these models for image reconstruction, quantitative tissue parameter extraction, and automatic optimization of acquisition protocols. In molecular MRI, however, the increased complexity of the imaging scenario, where the signals from various chemical compounds and multiple proton pools must be accounted for, results in exceedingly long model simulation times, severely hindering the progress of this approach and its dissemination for various clinical applications. Here, we show that a deep-learning-based system can capture the nonlinear relations embedded in the molecular MRI Bloch-McConnell model, enabling a rapid and accurate generation of biologically realistic synthetic data. The applicability of this simulated data for in-silico, in-vitro, and in-vivo imaging applications is then demonstrated for chemical exchange saturation transfer (CEST) and semisolid macromolecule magnetization transfer (MT) analysis and quantification. The proposed approach yielded 78%-99% acceleration in data synthesis time while retaining excellent agreement with the ground truth (Pearson's r>0.99, p<0.0001, normalized root mean square error < 3%).

**Introduction**

Mathematical modeling of biophysical effects has played a key role in the establishment, development, and routine clinical interpretation of magnetic resonance imaging (MRI)[1]. While proton spins, the primary source of signal in MRI, are governed by non-intuitive and complex quantum mechanics[2], their ensemble and net magnetization can be well described as a classical mechanic physics system, formulated as a set of coupled differential equations[3]. Based on these well-validated Bloch equations, imaging scientists were able to simulate offline the interactions between the tissue magnetic properties and the scanner-generated radio-frequency (RF) irradiation. This capability has enabled the development of new pulse sequences and image acquisition strategies aimed for enhancing a variety of clinically meaningful tissue characteristics[4,5].

Chemical exchange saturation transfer (CEST) MRI is an increasingly studied molecular imaging technique, capable of detecting millimolar concentrations of mobile proteins, peptides, and metabolites, via the exchange mechanism between the compound of interest labile protons and the water protons[6]. CEST imaging has been demonstrated as a promising tool for a variety of biomedical tasks including early stroke characterization[7], cancer detection and grading[8–10], kidney disease evaluation[11,12], reporter gene imaging[13–15], and metabolism characterization in neurological disorders[16]. Semisolid magnetization transfer (MT) imaging is another molecular MRI approach, where the data is acquired very similarly to CEST, yet it provides information on semisolid components, such as membranes or myelin sheets[17]. The semisolid MT contrast has been used for characterizing white-matter disease (such as multiple sclerosis), assessing the treatment response to cancer[18], and increasing the quality of magnetic resonance angiography (MRA) images[19].

Similar to conventional water-proton-based MRI, the progress of CEST and semisolid MT imaging was intertwined with the ability to formulate, explore, and solve the underlying mathematical model. Notably, such multi-proton-pool scenarios necessitate a substantial expansion of the Bloch model, which needs to account for the additional interplay between the exchangeable proton pools, the water pool, and the acquisition parameters. To meet these requirements, the expanded Bloch-McConnell (BM) model was developed[20] and later served as a powerful tool for a variety of molecular imaging applications, such as optimizing the design of paramagnetic CEST contrast agents[21], uncovering iopamidol's multi-site exchange properties and pH sensitivity[22], and the determination of the exchange parameters of various CEST compounds[23,24].

By fitting experimentally acquired data to the BM equations, researchers were able to reconstruct quantitative medical images and map clinically useful biophysical properties such as pH[25] and temperature[26]. While in steady-state CEST cases, the parameter fitting can be performed via analytical solutions of the BM equations[27], this approach mandates a lengthy acquisition, which requires the use of multiple saturation powers, potentially resulting in impractical scan times (e.g., more than an hour[28]). On the other hand, applying accurate BM fitting for a clinically relevant, non-steady-state, and faster acquisition scheme requires using the numerical model solution, leading to exceedingly long reconstruction times (on the order of hours[29]).

Recently, several deep-learning-based strategies were developed to accelerate the molecular MRI pipeline. CEST Z-spectrum data were converted into phosphocreatine concentration maps using a neural network (NN) trained with BM-simulated signals[29]. The extraction of amide proton volume fraction and exchange rates, as well as semisolid MT parameters, was successfully performed in-vivo, using pseudo-random MR-fingerprinting (MRF) acquisition schemes, empowered by deep reconstruction strategies (typically trained using numerical dictionaries containing millions of entries)[30–33].

However, all the above works rely on a computationally demanding generation of synthetic BM-based signals, requiring from hours to days, depending on the complexity of the imaging scenario addressed. This exhaustive and mandatory process must be repeated for every change in the acquisition protocol parameters or hardware available (e.g., modifying the recovery time, flip angle, duty cycle, or $B_0$ field strength), and for every new biological scenario of interest (e.g., where the metabolite of interest or relaxation properties may vary). This restriction severely limits the generalizability and utilization of quantitative CEST/semisolid MT techniques for the broad spectrum of molecular imaging applications. In addition, the lengthy BM simulation times hinder the efficient optimization of *any* semisolid MT/CEST acquisition protocols, as such efforts mandate an accurate simulation of a huge number of BM signals, which become impractical for the Guassian saturation pulse trains typically required for clinical applications and for super-Lorentzian lineshapes.

Here, we designed a deep-learning-based system that can capture the nonlinear relations embedded in the molecular MRI Bloch-McConnell model, enabling a rapid and accurate generation of biologically realistic synthetic data. The system was designed to allow robustness and flexibil-ity for various hardware and biological conditions while accommodating *any* acquisition protocol length. Validation was performed in-silico, in-vitro, and in-vivo for both classical CEST-weighted imaging and quantitative semisolid MT/CEST MRF.

Results

**Dynamic simulation of molecular MRI signals.** A fully connected NN was designed to receive tissue and scanner parameters of interest and rapidly generate the corresponding semisolid-MT or CEST signal (Fig. 1a). Each inference cycle was designed to output a single time-evolution signal element, which was then stored and served as input for the next cycle. This configuration enabled a synthesis unrestricted by acquisition protocol length. The accuracy of the rapid signal generator was initially examined in-silico, using six practical imaging scenarios[34], involving the following molecular targets: L-arginine, semisolid MT, aliphatic relayed nuclear Overhauser enhancement (rNOE), amine, iohexol, and amide. Representative pairs of NN-predicted and ground truth signals (obtained via traditional numerical solution of the BM equations) are shown in Fig. 2. An excellent agreement between the NN-generated and the ground-truth signals was observed for the entire test set (Fig. 3), with a normalized root-mean-squared error (NRMSE) smaller than 3%, and a significant correlation across all cases (Pearson's r>0.99, p<0.0001). The signal generation time was 78%-96% shorter compared to the reference consensus signal generator[31,35].

**In-vitro CEST MRF based on NN-generated signals.** To further validate the applicability of deep signal generation for standard imaging applications, an in-vitro phantom study was con-ducted. A 50 mM L-arginine phantom, composed of three vials titrated to different pH levels (yielding different proton exchange rates), was scanned at 9.4T. A dictionary of 665,873 simulated signals (Fig. 3 - top left), corresponding to various combinations of the water pool and the amine proton exchange parameters[36,37] was generated using the NN-based and the standard reference method. A traditional dot-product-based MRF reconstruction was then performed to quantify the L-arginine properties using the two dictionaries. The resulting proton exchange rate (Fig. 4b,f) and concentration (Fig. 4a,e) maps obtained using both methods were in excellent agreement, demonstrating an NRMSE<2% and a structural similarity index (SSIM)>0.90.

**In vivo semisolid MT MRF based on NN-generated data.** As another means to assess the NN-based signal generator under realistic settings, a semisolid MT brain imaging experiment was performed in a wild-type mouse scanned at 9.4T. A dictionary composed of 26,400 signals representing various tissue parameter combinations was generated using the same dynamic NN used in the L-arginine phantom study. The resulting dictionary was used for dot-product MRF reconstruction of the semisolid MT proton volume fraction and exchange rates across the mouse's brain and compared to the corresponding images reconstructed using a traditional dictionary synthesis. The resulting exchange parameter maps generated using both methods (Fig. 4c,d,g,h) were in excellent agreement, demonstrating an NRMSE<4% and a structural similarity index (SSIM)>0.98.

**Application optimized ultrafast signal generation.** For applications focused on a single exchangeable proton pool or a particular pathology of interest, the optimization of the signal-to-noise ratio, or parameter quantification ability is highly desirable, but the flexibility to accommodate a variety of imaging cases and various acquisition schedule lengths can be relaxed. To accommodate these situations, we have designed another NN-based molecular signal generation architecture (Fig. 1b), where the dynamic component is absent, and the entire signal trajectory (including all its time-evolution components) is generated simultaneously. While such architecture can only enable a fixed-size signal generation, it inherently offers increased acceleration compared to the dynamic NN (Fig. 1a), requiring only a single inference cycle. The system was validated using a challenging imaging scenario, where CEST signals from the human brain need to be synthesized while considering the simultaneous effects from seven proton pools (amide, guanidine, amine, OH, NOE, semisolid MT, and water)[17]. The simulation was performed using a clinically relevant saturation pulse train, which is particularly lengthy to simulate using standard signal generators. A comparison between a representative NN-generated Z-spectrum from the above-described case and its traditionally simulated counterpart is shown in Fig. 5a. The excellent agreement between the entire test-set results obtained with both methods is shown in Fig. 5b (Pearson's r>0.99, p-value<0.0001, NRMSE<1%). Notably, the test-set generation time was shortened from almost 24h using the reference method to 1.2s using the NN (72,000 fold acceleration).

**Discussion**

The BM equations have played a vital role in the progress of semisolid MT and CEST imaging. Their solution provides offline insights on protocol design, field strength dependencies[38], and predicted compound-related effects, and is being routinely used in a variety of imaging tasks, such as traditional Bloch fitting[39] and semisolid-MT/CEST MRF[32]. While various deep-learning-based strategies were recently proposed for accelerating/quantifying molecular semisolid MT/CEST imaging [29,30,34,40–44], most of these computational approaches require an exhaustive simulation of the underlying BM model that needs to be repeated for millions of different tissue parameter combinations. The proposed NN-based signal generator is expected to complement these quantitative CEST efforts, providing a drastic acceleration at a different point in the imaging pipeline - the dictionary generation step. Moreover, the same signal generator could be used for accelerating classical Bloch-fitting of traditional CEST data. Notably, the dynamic nature of the suggested framework (Fig. 1a) offers a robust framework for accommodating *any* signal length and a wide variety of tissue and scanner parameters while circumventing the need to re-synthesize data for a newly examined imaging protocol and biological scenario. This characteristic is expected to enable rapid comparison between the encoding capability of various-length quantitative CEST protocols, assist in choosing the optimal field strength for a given application, and improve the contrast to noise-ratio in semisolid MT or CEST-weighted imaging.

While analytical solutions for semisolid MT and CEST-weighted signals were previously derived [27,38,45,46], they rely on several inherent assumptions that yield inaccuracy at particular exchange regimes and are incompatible with many saturation pulse trains and pulse shapes. As numerically solving the BM equations in such cases is extremely computationally demanding, the deep signal generator becomes particularly attractive, as its inference time is agnostic to the pulse shape, saturation pulse train characteristics, and the macromolecular Lorentzian/super-Lorentzian absorption lineshape. Specifically, a 72,000-fold acceleration was demonstrated here (Fig. 5) to simulate a seven-pool CEST Z-spectrum acquisition using a sinc-Gaussian pulse trains.

The choice of the fully connected architecture for the deep signal generator was based on its previous success in capturing semisolid MT/CEST signal dynamics[30] and as a means to minimize the model complexity, training, and inference time. Future work could nonetheless explore more sophisticated recurrent-based architectures[47], which may further improve the system's accuracy (though with a potential penalty in computational time).

A single dynamic model (with a fixed set of optimized neural network weights) was able to accurately generate semisolid MT/CEST signals (NRMSE<3%, Pearson's r>0.99, p<0.0001) for six different imaging scenarios (Fig. 2 and Fig. 3). The same model could be expanded and trained to accommodate additional pulse-sequence diagrams and various readout patterns used by the MRI community. This process is expected to be relatively straightforward via the use of the open-source pulseq standard format sequences[31]. Similarly, the same model could be used to generate traditional water pool $T_1$ and $T_2$ signal dictionaries for conventional MRF[48] while taking into consideration the magnetization transfer effect on the measured relaxivity[49,50].

As demonstrated in Fig. 3 and Fig. 5b, there is a tradeoff between accuracy and inference time to flexibility. This is mainly expressed in the increased correlation between the NN-predicted and ground truth in the application-optimized network compared to the dynamic network. While use of the generalized dynamic network enabled satisfying reconstruction accuracy for multiple imaging scenarios when examined on "real" experimental data (Fig. 4), a researcher/physician with a research focus on a particular imaging application (e.g., amide proton transfer brain imaging) may opt to use the application optimized approach, as it provides even higher accuracy.

## Conclusion

A deep learning-based framework for dynamic and rapid generation of semisolid molecular MT and CEST MRI signals was developed, demonstrating broad applicability with various imaging scenarios and acquisition protocols.

## Methods

**Deep molecular signal generator architecture.** Two deep learning models were designed to receive a set input of tissue and scan parameters and rapidly generate the corresponding semisolid MT or CEST signals. Both architectures consisted of a four-layers fully connected neural network, with 256x256 neurons in the two hidden layers and sigmoid activation functions. The *dynamic* network variant (Fig. 1a) was designed to accommodate a two-pool imaging scenario of various target compounds using a continuous wave saturation pulse scheme with a wide range of saturation and readout characteristics (Table 1), and a variety of field strengths ($B_0$ = 3, 4.7, 7, 9.4, and 11.7). The network operated in an iterative manner, where each inference cycle *i* yielded a single time-evolution signal element ($S_i = |M_{xy_i}|$). In the next cycle, the last estimated signal served as additional input for generating $S_{i+1}$ until reaching the required signal length N. Training was performed using 104,190,000 random signal trajectories from the tissue and scanner parameter range described in Table 1. The data was split between training and validation sets in a ratio of 90/10%, employing an early-stopping approach to prevent over-fitting. A separate set of 1,694,183 signals from various imaging scenarios[34] was synthesized as a test set.

The *application-optimized* network variant (Fig. 1b) was designed to obtain a further acceleration in signal generation time, with the cost of removing the flexibility for variable output signal acquisition lengths. The network was oriented to address a seven-proton-pool human brain imaging scenario at a clinical scanner, with various tissue parameter values enabled, based on the baseline properties described in [17,51]. A flexible CEST protocol, consisting of Sinc-Gaussian pulses, variable pulse duration, variable number of saturation pulses, recovery time, and $B_0$ magnetic field was considered, and the output consisted of an N=34 long acquisition schedule, obtained in a single inference cycle. The training was performed using 5,443,200 random signal trajectories from the tissue and scanner parameter range described in Table 2, with a 90/10% split between training and validation data. A separate set of 200,000 signal trajectories was synthesized for the test phase.

For both network variants, the training was performed using signal dictionaries generated via a numerical reference solution of the Bloch-McConnell equations (see next section), employing the adaptive moment estimation (ADAM) optimizer, a mean square error loss, a learning rate of 0.0001, and a batch size of 2048. The NNs were realized using the TensorFlow framework, a GeForce RTX 3060 TI graphic processor, and a 12-core CPU desktop. The total training time was 640 min and 35 min for the dynamic and application-optimized network variants, respectively.

**Reference standard molecular signal generator.** Dictionaries of simulated signal intensity trajectories were generated using a state-of-the-art Bloch-McConnell equations numerical solver[31,35], implemented in $C^{++}$ with a SWIG wrapper for Python, according to the open-source pulseq-based standard[52,53]. To provide a fair comparison for the MRF application, requiring the simultaneous and exhaustive generation of a huge number of signal trajectories, we have activated the pulseq-signal generator in a parallel execution manner (see the code availability section).

**CEST phantom imaging.** An in-vitro MRF imaging study was performed using an L-arginine phantom composed of three different vials with a 50 mM concentration and a pH titrated to 5.0, 5.5, and 6.0 pH (affecting the amine proton exchange rate)[36,37]. Imaging was conducted using a 9.4T MRI scanner (Bruker Biospin, Billerica, MA), a transmit/receive volume coil (Bruker Biospin, Billerica, MA), a field of view (FOV) of 32x32 mm$^2$, a matrix of 64x64 pixels, and a 5 mm slice thickness. An in-house programmed, flexible semisolid MT/CEST-EPI protocol was used, employing a pseudo-random varied series of saturation pulse parameters, as described in[36,37].

**In vivo semisolid MT mouse imaging.** All animal experiments and procedures were performed in accordance with the NIH Guide for the Care and Use of Laboratory Animals and were approved by the Institutional Animal Care and Use Committee of the Massachusetts General Hospital. A C57/BL6 wild-type male mouse was purchased from Jackson Laboratory. It was anesthetized using 1%–2% isoflurane and placed on an MRI cradle with ear and bite bars to secure the head. The respiration rate was monitored with a small animal physiological monitoring system (SA Instruments, Stony Brook, NY), and the temperature was maintained by blowing warm air in the bore of the magnet. A quadrature volume coil was used for RF transmission, and a mouse brain phased array surface coil was used for receive (Bruker Biospin, Billerica, MA). A field of view (FOV) of 19x19 mm$^2$, a matrix of 64x64 pixels, and a 1 mm slice thickness were used. A pseudo-random semisolid-MT MRF protocol was implemented, where the saturation pulse power and frequency offset were varied between differently acquired raw images, as described in[30].

**Statistical analysis.** Pearson's correlation coefficients were calculated using the open-source SciPy scientific computing library for Python[54]. The structural similarity index (SSIM) was computed using the SSIM-python imaging library (PIL). Differences were considered significant at $p<0.05$.

**Data availability.** The datasets used in this study are publicly available at: [an updated link will be inserted upon acceptence].

**Code availability.** Open-source code is available at: [an updated link will be inserted upon acceptence].

**Acknowledgments**

This work was supported by the Ministry of Innovation, Science and Technology, Israel and a grant from the Tel Aviv University Center for AI and Data Science (TAD).

**Author contributions**

D.N. and O.P. conceptualized the problem. D.N., C.T.F., and O.P. contributed to experimental design. C.T.F. and O.P. acquired the imaging data. D.N., N.V., and O.P. designed and/or implemented the signal generators. D.N., N.V., and O.P. analyzed the results. D.N., N.V., C.T.F., and O.P. wrote and/or substantially revised the manuscript.

**Additional Information**

**Competing Interests:** The authors declare no competing interests.

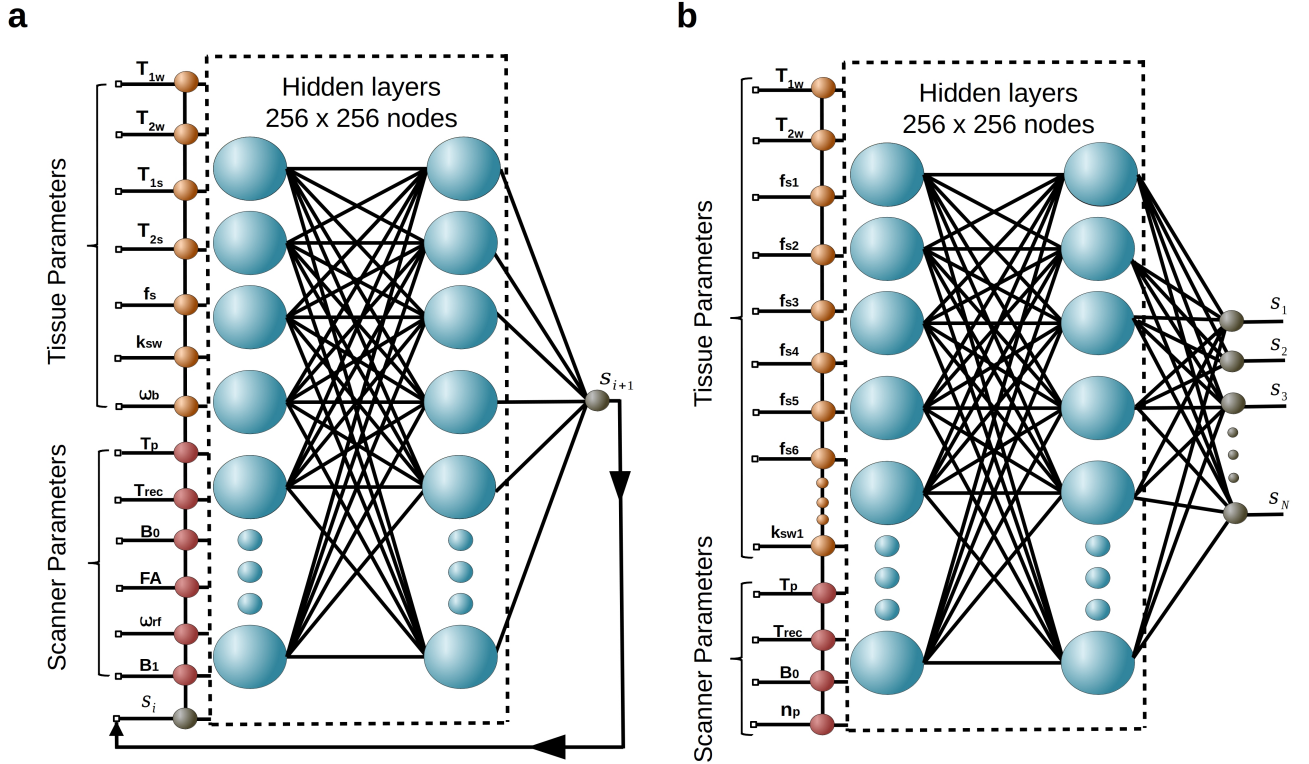

**Fig. 1. Network architectures. a.** A dynamic fully connected NN receives tissue parameters, scanner parameters, and the previous inference cycle signal $S_i$. The output is the next time-evolution signal element $S_{i+1}$. The inference cycle continues according to the user's desired signal acquisition length N. **b.** Application Optimized network. The input of the network are tissue and scanner parameters, and the output is the entire magnetic signal inferred at once, according to a particular value N used during the training. $T_{1w}$ - water longitudinal relaxation, $T_{2w}$ - water transverse relaxation, $T_{1s}$ - solute longitudinal relaxation, $T_{2s}$ - solute transverse relaxation, $f_{si}$ - proton volume fraction for solute i, $k_{swi}$ - proton exchange rate for solute i, $\omega_b$ - chemical shift, $T_p$ - saturation pulse duration, $T_{rec}$ - recovery time, $B_0$ - main magnetic field, FA - flip angle, $\omega_{rf}$ - saturation pulse frequency offset, $B_1$ - saturation pulse power, $n_p$ - number of pulses in the saturation pulse train.

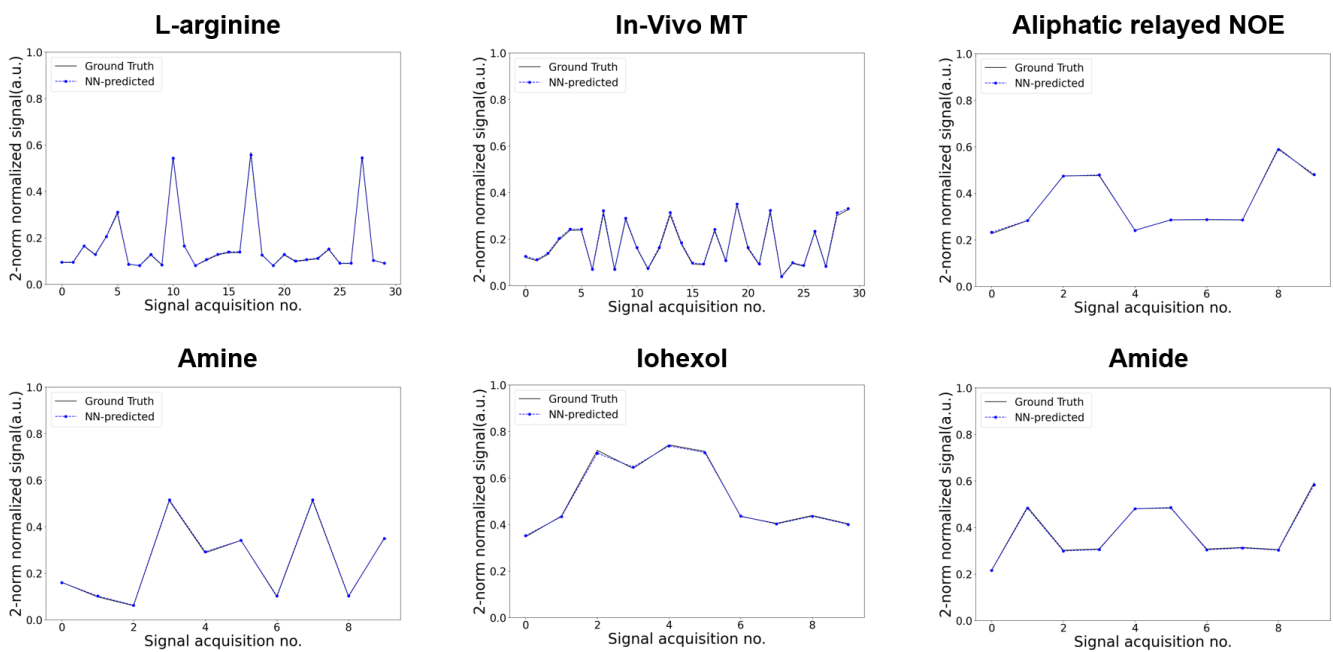

**Fig. 2. In-silico comparison between dynamic-NN-generated semisolid MT/CEST signals, and their ground-truth counterparts**. Each subfigure describes a representative test-set signal trajectory taken from a different molecular imaging scenario[34], all generated using a single NN (Fig. 1a). Note the excellent agreement between the reference ground-truth standard and the NN-generated trajectories.

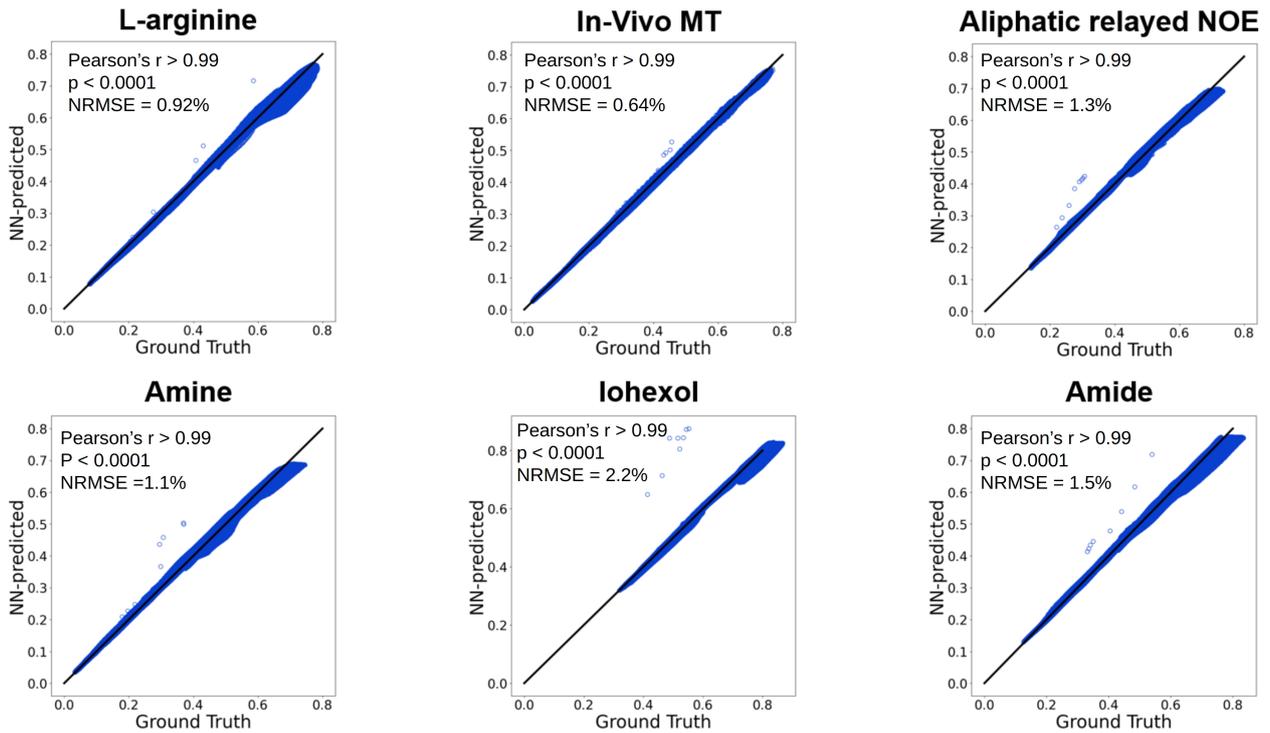

**Fig. 3. Statistical analysis of the in-silico signal generation experiment performed using the dynamic NN.** An excellent agreement was obtained between the NN-predicted signal trajectories and their ground truth counterparts (Pearson's r>0.99, p<0.0001, normalized root mean square error (NRMSE)<3%) for all six examined imaging scenarios.

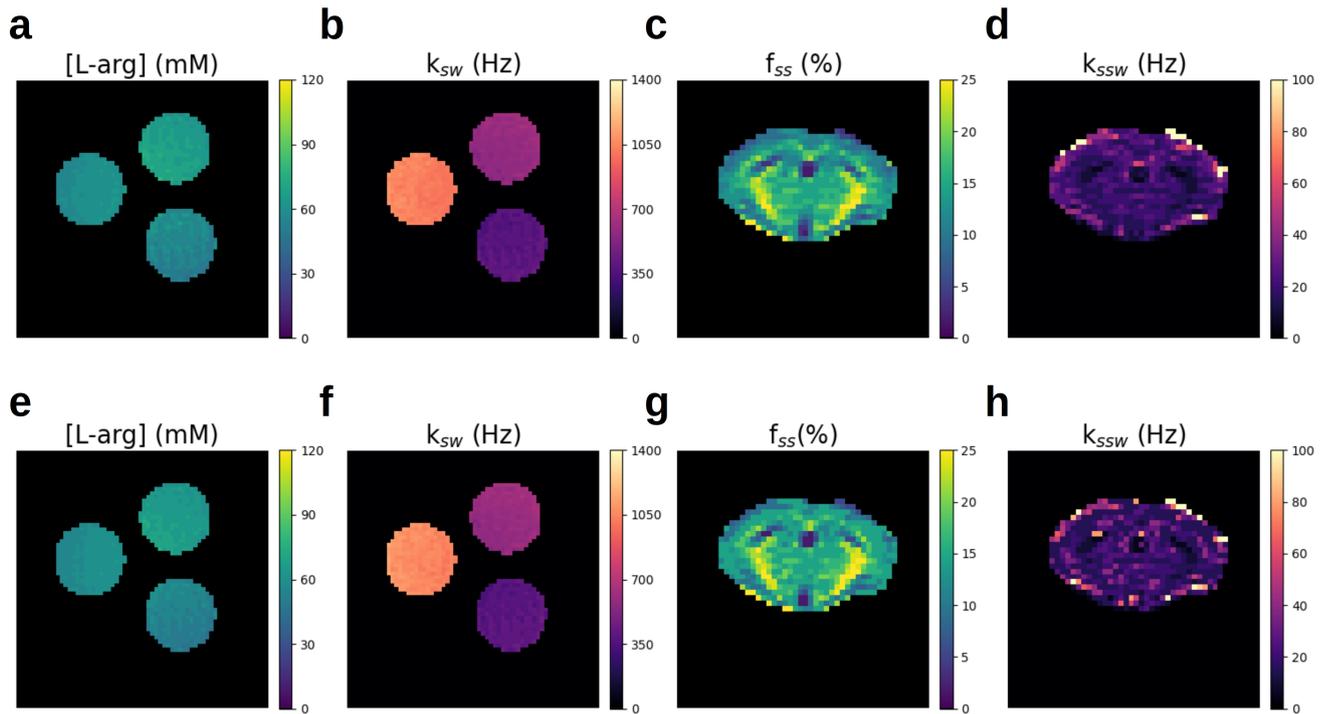

**Fig. 4. Magnetic resonance fingerprinting quantitative reconstruction of in-vitro CEST and in-vivo semisolid MT data using dynamic-NN-generated signals. a-b, e-f**. L-arginine concentration and proton exchange rate maps were obtained using dynamic-NN-generated signals (top) and traditional numerical solutions of the BM equations (bottom). **c-d, g-h**. In vivo semisolid MT proton volume fraction and exchange rate maps from a wild-type mouse, obtained using dynamic-NN-based signals (top) and traditional numerical solution of the BM equations (bottom).

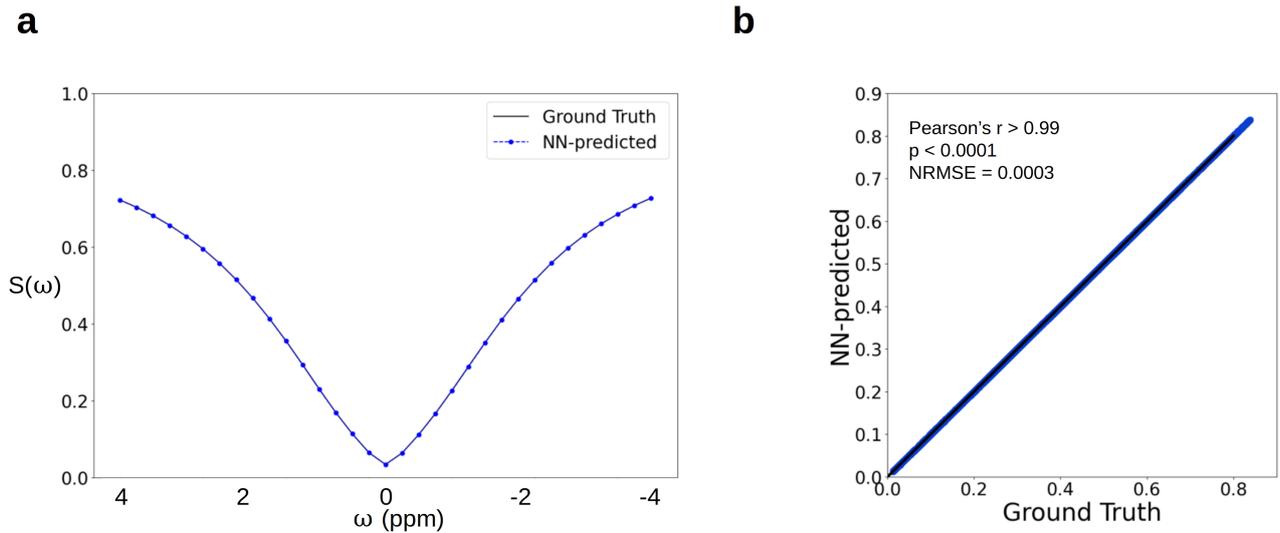

**Fig. 5. In-silico generation of a seven-pool human brain Z-spectrum data, acquired with a Sinc-Gaussian saturation pulse train, using an application optimized NN. a**. A representative brain CEST signal generated using the NN-predicted (blue) or the reference standard method (black). **b**. Statistical analysis of the entire test set data, demonstrating an excellent agreement between the NN-predicted and ground truth reference values (Pearson's r>0.99, p<0.0001, normalized root mean square error = 0.0003).

**Table 1.** Tissue and scanner parameter range used for dynamic NN training

| Parameter | Minimum value | Maximum value |
|---|---|---|
| $T_{1w}$ (s) | 1.3 | 3.4 |
| $T_{2w}$ (s) | 0.04 | 1.2 |
| $T_{1s}$ (s) | 1.3 | 3.4 |
| $T_{2s}$ (s) | $4 \cdot 10^{-5}$ | 0.04 |
| $f_s$ | $\frac{20}{110000}$ | $\frac{30000}{110000}$ |
| $k_{sw}$ (s$^{-1}$) | 5 | 1500 |
| $B_1$ ($\mu T$) | 0.25 | 6 |
| $T_p$ (s) | 1 | 10 |
| $T_{rec}$ (s) | 1 | 10 |
| $B_0$ (T) | 3 | 11.7 |
| FA (°) | 60 | 90 |
| $\omega_b$ (ppm) | -3.5 | 4.3 |
| $\omega_{rf}$ (ppm) | -3.5 | 20 |

Notations are defined in the caption of Fig.1

**Table 2.** Tissue and scanner parameter range used for application-optimized NN training

| Parameter | Minimum value | Maximum value |
|:---:|:---:|:---:|
| $T_{1w}$ (s) | 0.8 | 1.6 |
| $T_{2w}$ (s) | 0.04 | 0.12 |
| $f_s$ amide | 0.0005 | 0.0045 |
| $k_{sw}$ amide (s$^{-1}$) | 30 | 90 |
| $f_s$ guanidine | 0.0005 | 0.0014 |
| $f_s$ amine | 0.0005 | 0.0045 |
| $f_s$ OH | 0.0005 | 0.0045 |
| $f_s$ NOE | 0.0022 | 0.0067 |
| $f_{ss}$ MT (Lorentzian) | 0.018 | 0.216 |
| $B_1$ ($\mu T$) | 0 | 6 |
| $T_p$ (s) | 0.05 | 0.1 |
| $T_{rec}$ (s) | 2.5 | 4 |
| $B_0$ (T) | 3 | 11.7 |
| $n_p$ | 10 | 30 |

Notations are defined in the caption of Fig.1. The fixed tissue parameters are available at the pulseq-CEST simulation repository[51], based on the values described in[17]. The fixed acquisition protocol parameters are defined at the pulseq sequence library[55], as described in the CEST APT consensus paper, sequence APTw002[35].